\documentclass[12pt,preprint,numeredappendix]{aastex}
\usepackage{epsf}

\newcommand{\tx}[1]{\textrm{#1}}

\newcommand{\ltsima}{$\buildrel<\over\sim$}
\newcommand{\lapprox}{\lower.5ex\hbox{\ltsima}}
\newcommand{\kms}{km~$\tx{s}^{-1}$}
\newcommand{\dv}{$r^{1/4}\,$}

\newcommand{\be}{\begin{equation}}
\newcommand{\ee}{\end{equation}}

\newenvironment{inlinefigure}{
\def\@captype{figure}
\noindent\begin{minipage}{0.999\linewidth}\begin{center}}
{\end{center}\end{minipage}\smallskip}

\slugcomment{ApJ, in press}

\shorttitle{Keck Spectroscopy of GOODS Spheroidals}
\shortauthors{Treu et al.}

\begin{document}

\title{Keck Spectroscopy of distant GOODS Spheroidal Galaxies: Downsizing in a Hierarchical Universe}

\author{Tommaso Treu\altaffilmark{1,2}, Richard
S. Ellis\altaffilmark{2}, Ting X. Liao\altaffilmark{2}, Pieter G. van
Dokkum \altaffilmark{3,2}}

\altaffiltext{1}{Department of Physics and Astronomy,
University of California at Los Angeles, Los Angeles, CA 90095; ttreu@astro.ucla.edu; Hubble Fellow}
\altaffiltext{2}{Caltech, Astronomy, 105-24, Pasadena, CA 91125,
USA; rse@astro.caltech.edu, ting@caltech.edu}
\altaffiltext{3}{Department of Astronomy, Yale University, New
Haven, CT 06520; dokkum@astro.yale.edu}


\begin{abstract}
We analyze the evolution of the Fundamental Plane for 141 field
spheroidal galaxies in the redshift range 0.2$<$$z$$<$1.2, selected
morphologically to a magnitude limit F850LP$=$22.43 in the northern
field of the Great Observatories Origin Survey. For massive galaxies
we find that the bulk of the star formation was completed prior to
$z=2$. However, for the lower mass galaxies, the luminosity-weighted
ages are significantly younger. The differential change in
mass-to-light ratio correlates closely with rest-frame color,
consistent with recent star formation and associated growth. Our data
are consistent with {\it mass} rather than {\it environment} governing
the overall growth, contrary to the expectations of hierarchical
assembly.  We discuss how feedback, conduction, and galaxy
interactions may explain the downsizing trends seen within our large
sample.
\end{abstract}

\keywords{galaxies: elliptical and lenticular, cD --- galaxies:
evolution ---- galaxies: formation --- galaxies: structure --
cosmology: observations}

\section{Introduction}

The question of the assembly history of elliptical and S0 (hereafter
spheroidal) galaxies has come into sharp focus in recent years with
the suggested discovery of their possible progenitors at redshifts
above 2 (e.g., Chapman et al.\ 2003). However, the topic remains
controversial both from a theoretical and observational standpoint.

Semi-analytic models based on the current $\Lambda$CDM structure
formation picture predict for spheroidals remarkably different
assembly rates than observed (Benson et al. 2002), suggesting the
effects of merging on stellar mass assembly are poorly
understood. Consequently, the implication of recent claims for an
abundant population of old massive systems at $z\simeq$2 are unclear
(Glazebrook et al.\ 2004; Cimatti et al.\ 2004).

Exactly how the slow evolutionary trends observed in cluster
spheroidals (Kelson et al. 1997; Bender et al. 1998; Stanford,
Eisenhardt \& Dickinson 1998; van Dokkum \& Stanford 2003) contrast
with those measured in the field (Treu et al 1999,2001b, 2002; van
Dokkum \& Ellis 2003; Gebhardt et al.\ 2003; Rusin et al.\ 2003;
Bernardi et al.\ 2003; Yi et al.\ 2005; Jimenez et al.\ 2004) is also
unclear. The environmental dependencies do not appear to be as large
(van Dokkum et al. 2001) as predicted by some renditions of
hierarchical models (e.g. Diaferio et al.\ 2001; de Lucia et al.\
2004). And those trends indicative of early assembly appear to be
strongly at variance with the rapid decline in the red luminosity
density claimed from large photometric samples (Bell et al 2004).

In this Letter we present the first results from a comprehensive Keck
survey of 141 field spheroidals selected in the northern Great
Observatories Origins Survey (GOODS) field. The spectroscopic signal
to noise enables us to measure accurate stellar velocity dispersions
for each spheroidal and, utilizing the Fundamental Plane (FP,
Djorgovski \& Davis 1987; Dressler et al.~1987), the dynamical mass
and mass-to-light ratio. As these masses are closely linked to those
of the underlying halos (Treu \& Koopmans 2004), we establish a more
complete picture of the assembly history than from studies based
solely on photometrically-derived stellar masses. Our sample achieves
a high degree of statistical completeness and is almost an order of
magnitude larger than any previous equivalent study. This enables us
to examine the rate of assembly as a function of dynamical mass and
environmental density, both key variables for testing the
currently-popular hierarchical picture for galaxy formation.

We assume throughout a cosmological model with $\Omega_{\rm M}=0.3$,
$\Omega_\Lambda=0.7$ and H$_0$=65 km~s$^{-1}$~Mpc$^{-1}$. All
magnitudes are in the AB system (Oke 1974), unless otherwise noted.

\section{Data}

\label{sec:sampobs}

In order to compare properly with local E+S0 galaxies, our parent
sample was selected morphologically using visual classification to a
F850LP magnitude limit of 22.43 (AB). Internal tests (Treu et
al. 2005, in preparation, hereafter T05) show that our classification
is self-consistent. The surface density of spheroidals is in agreement
with that found by comparable studies (e.g. 1.32$\pm$0.09
arcmin$^{-2}$ vs. 1.23$\pm$0.10 arcmin$^{-2}$ measured by Im et
al. 2002 to the same depth), showing that our classification is
consistent with previous work and that the surface density is
comparable with that expected from the local luminosity function.

High signal-to-noise spectra were obtained during April 1-5 2003 using
the DEIMOS spectrograph at the Keck-II Telescope. Conditions were
excellent with seeing in the range $0\farcs6$-$0\farcs8$
throughout. The 1200 grating blazed at 7500\AA\, was centered at 8000
\AA.  Total exposure times ranged between 14,400 and 38,900
seconds. The DEIMOS spectra were reduced using the IDL pipeline
developed for the DEEP2 Survey (Davis et al 2003).

Stellar velocity dispersion were derived for 141 spheroidals (out of
163 spectroscopically targeted) using the methods described in Treu et
al.\ (1999,2001a). To estimate systematic errors, a subsample was
analyzed independently using methods described in van Dokkum \& Franx
(1996). The independent determinations agree within 1\% on average
(12\% rms), indicating that systematic differences due to methodology
are negligible. Comparisons with previously-published dispersions in
the Hubble Deep Field (van Dokkum \& Ellis 2003) as well as repeat
DEIMOS measures for targets on different masks show that our estimated
errors ($\sim10\%$) are reliable. Surface photometry through \dv\,
model fitting was derived from the images using the GALFIT package
(Peng et al.\ 2002).

A comparison of the subsample with measured velocity dispersions with
the parent sample of morphologically selected spheroidals shows it to
be representative with respect to color, luminosity and redshift
distribution. We conclude that our sample is representative of
morphologically selected spheroidals.  The catalog, a description of
the sample, related selection effects, observations and data reduction
are given in T05.


\label{ssec:ld}

In order to test for environmental trends, local projected densities
were derived for individual spheroidals using a modification of the
algorithm described in Dressler (1980) and the GOODS-N photometric
catalog of Bundy, Ellis \& Conselice (2004; hereafter B04).  The B04
catalog includes spectroscopic redshifts from this survey and the Keck
Team Redshift Survey (Wirth et al.\ 2004), with relatively uniform
spatial sampling, supplemented by photometric redshifts (for the
remaining 45\% of the objects).  For each spheroidal, we identify
nearby objects brighter than M$_{\rm V}=-19.8$, allowing for passive
evolution. Then, moving away on the plane of the sky, we add the
probability that each neighbor lies within 3000 \kms\, until a total
number 5 is reached. This defines the local density $\Sigma_5$.  We
chose 5 nearest neighbors (rather than the usual 10) to enhance
sensitivity to more modest overdensities, resulting in a poisson noise
component of $\sim$0.2 dex. The partial use of photo-z brings the
total error on $\Sigma_5$ to 0.3 dex per galaxy (as calculated with
simulations) .

\section{Results}

\subsection{Evolution of the Fundamental Plane}

The Fundamental Plane (FP) is defined as:
\begin{equation}
\label{eq:FP} \log R_{\tx{e}} = \alpha \log~\sigma + \beta~\tx{SB}_{\tx{e}} +
\gamma.
\end{equation}
where R$_{\tx{e}}$ and SB$_{\tx{e}}$ are the effective radius and
surface brightness and $\sigma$ is the central velocity dispersion. A
physical interpretation of the evolution of the FP follows from defining 
an effective (dynamical) mass,
\be
M \equiv \frac{5\sigma^2 R_{\tx{e}}}{G}.
\label{eq:mass}
\ee
Redshift variations of the slopes $\alpha$, $\beta$ and intercept
$\gamma$ can be interpreted via evolution of the stellar
populations. If $\sigma$ and R$_{\tx{e}}$ and Eq.~\ref{eq:mass} do
not evolve, for the $i$th galaxy:
\be
\gamma^i\equiv \log R_{\tx{e}}^i - \alpha \log \sigma^i - \beta \tx{SB}_{\tx{e}}^i,
\label{eq:gammai}
\ee
and the offset from the FP ($\Delta \gamma^i \equiv \gamma^i-\gamma$)
is related to the offset of the $M/L$ by
\be
\Delta \log \left( \frac {M}{L} \right)^i = -\frac{\Delta \gamma^i}{2.5 \beta}.
\label{eq:dgdml}
\ee
The scatter of the FP ($\sim0.1$ in $\log R_{\rm e}$; Bernardi et al.\
2003) is thus a measure of the homogeneity of the stellar population
(see Treu et al.\ 2001b for more discussion).

Figure~\ref{fig:FPBp4} shows the offset from the relation for the
local Coma Cluster\footnote{We use Coma as the local reference for
both cluster and field to minimize systematic uncertainties related to
filter transformations, distance determination and selection effects;
see discussion in Treu et al.\ 2001b.}  using Eq.~\ref{eq:dgdml} for
all 141 GOODS-N spheroids coded by dynamical mass
(Eq~\ref{eq:mass}). Evolutionary tracks for single burst stellar
populations (Treu et al. 2001b) at various formation redshift ($z_{\rm
f}=1,2,5$) are shown for comparison.  The most massive spheroidals
uniformly follow a passive evolutionary trend consistent with an early
epoch of formation, as found for massive cluster galaxies (e.g. van
Dokkum \& Stanford 2003).  However, those with smaller masses are
systematically younger (as hinted previously by smaller samples; Treu
et al.\ 2002, van der Wel et al.\ 2004) and show increased scatter,
implying a lower redshift of formation or secondary episodes of
activity which rejuvenated a previously old population (e.g. Trager et
al.  2000; Treu et al. 2002; van Dokkum \& Ellis 2003).

We can verify that recent star formation is responsible for the trends
in Figure~\ref{fig:FPBp4} by comparing $M/L$ with independent
diagnostics of recent stellar formation, such as the rest-frame color
(Figure~\ref{fig:DFPBV}). If we assume that the stellar $M_*/L_B$ can
be obtained as $\Delta \log M/L_B$ + $\log M_*/L_{\rm B, z=0}$ with
$M_*/L_{\rm B,z=0}=7.3 M_{\odot}/L_{\rm B,\odot}$ (Treu \& Koopmans
2004), we can reproduce the locus for an ageing single burst
population with a Salpeter IMF (red line in
Figure~\ref{fig:FPBp4}). The observed correlation is fairly close to
this model, although the scatter is significant, presumably as a
result of a mixture of primary and secondary episodes of star
formation, and, possibly, of metallicities.  The analysis of spectral
diagnostics such as Balmer absorption lines (T05) is also consistent
with the trend in $M/L$ being due to recent starformation.

\subsection{The role of mass and environment}

Our large and homogeneous sample allows us to take a significant step
forward with respect to previous studies by exploring further the
trends of M/L with dynamical mass introduced in \S 3.1 and by
examining the correlations between M/L and local galaxy density,
$\Sigma_5$.  Dynamical mass can be an effective proxy for the total
halo mass (Treu \& Koopmans 2004) and local density (\S~\ref{ssec:ld})
for environmental effects. Therefore, we can use these correlations to
study the role of halo mass and environment in the context of
contemporary models of structure formation. To this aim, we will
consider the quantity $\delta \Delta \log M/L_B$, defined as the
difference between $\Delta \log M/L_B$ for an individual spheroidal
and that expected for old massive cluster galaxies at that redshift
(i.e. $\Delta \log M/L_B=-0.46 z$, van Dokkum \& Stanford 2003).
Negative values of $\delta \Delta \log M/L_B$ indicate a smaller
$M/L_B$ than that of a massive cluster spheroid.

Figure~\ref{fig:DelDel4} shows $\delta \Delta \log M/L_B$ as a
function of galaxy mass; a clear trend is evident with lower
mass galaxies displaying younger luminosity-weighted ages. Although
the two axes are not independent (a larger mass implies a larger $M/L$
neglecting errors on $L_B$), both the small error bars (typically
$<0.1$ dex) and correlation with independent diagnostics
(Figure~\ref{fig:DFPBV}; T05) indicate this trend does not arise
from correlated uncertainties.  Because of age and luminosity limits,
a parent mass-selected sample must generally occupy a triangular
region between the horizontal blue line and the lowest observed
$\delta \Delta \log M/L_B$ at any given mass.  We thus conclude that
the luminosity-weighted age of spheroidals increases with dynamical
mass, a result we statistically confirm using detailed Monte Carlo
simulations in T05.

We illustrate the effects of secondary bursts of star formation in
Figure~\ref{fig:DelDel4} by considering a family of two burst models,
arranged by adding a young stellar component ( aged 1 Gyr) to an
earlier population (10 Gyr) equal for simplicity to the dynamical
mass. For example a secondary burst of $10^{9}$-$10^{10} M_{\odot}$
would reproduce the observed trend.

Figure~\ref{fig:DDS} shows $\delta \Delta \log M/L_B$ as a function of
local density $\Sigma_5$. No statistically significant correlation is
present, although the dynamic range in GOODS-N does not include
massive clusters. Within our larger error bars, our data are
consistent with the trend observed locally in the Sloan Digital Sky
Survey (SDSS; Bernardi et al.\ 2003), which in turn can be explained
in terms of mass differences in the context of environment-dependent
mass functions.  More pronounced is the difference with massive
galaxies in the cores of rich clusters ($\delta\Delta \log
M/L_B\equiv0$), although the poor statistics available for clusters at
comparable redshifts and the different selection criteria prevent a
detailed comparison.

\section{Discussion}

We have shown that mass is a key variable determining the star
formation history of field spheroidals. The most massive examples
($>10^{11.5}$ M$_{\odot}$) formed most of their stars before
$z\simeq\,2$, while less massive ones ($<10^{11}$ M$_{\odot}$) show
greater scatter and systematically younger stellar ages. The secondary
activity necessary to reproduce our observations (up to $\sim10^{10}$
M$_{\odot}$ over the past few Gyr) could be explained by accretion of
satellites or gas-rich mergers.  Regardless of the process,
substantial recent growth is implied in many cases for galaxies less
massive than 10$^{11}$ M$_{\odot}$.

Such ``downsizing'' (i.e. stellar growth that moves systematically to
lower mass systems as the Universe expands; Cowie et al. 1996) appears
contrary to the expectations of hierarchical models of galaxy
formation (Diaferio et al. 2001). In this widely-discussed picture,
massive spheroidals assembled recently via major mergers with
associated star formation. Strong environmental trends are also
expected in such a picture (e.g. De Lucia et al.\ 2004), as a result
of the fact that evolution is accelerated in these ``biased''
regions. We do not find any environmental trend within the range of
densities covered by the GOODS-N field, although a larger sample of
data will be needed to confirm this and properly compare with dense
clusters, by studying the bivariate distribution (mass and local
density) of $\delta \Delta \log M/L_{\rm B}$.

Our observations may be reconciled with the hierarchical paradigm via
the introduction of more detailed physics that regulates star
formation in collapsing systems. If, above a certain halo mass,
feedback from Supernovae or AGN, or thermal conduction (Benson et al.\
2003; Nagashima et al.\ 2004; Granato et al.\ 2004), heat or remove
gas and prevent it from cooling back onto the galaxy -- thus
inhibiting star formation -- then the most massive galaxies will
appear old even if they assembled dynamically at recent times,
e.g. via collisionless mergers.  Alternatively, star formation could
be anticipated in massive system, triggered by galaxy interactions
(Menci et al.\ 2004). Although infalling satellites or minor mergers
may be frequent, they are likely to influence only the lower mass
systems, as observed. The extended timescale necessary for lower mass
mergers to relax and present smooth morphological systems may also
play a key role (Hernandez \& Lee 2004). Regardless of the detailed
explanation, it is clear that the mass-dependent growth of spheroidals
is more complex than previously thought. It will be interesting to see
if the numerical models with these additional mechanisms can reproduce
quantitatively our result.

\acknowledgments

We acknowledge the use of the Gauss-Hermite Pixel Fitting Software
developed by R.~P.~van der Marel and financial support by NASA (Hubble
Fellowship HF-01167.01; STScI-AR-09960) and NSF (AST-0307859). We
thank those who developed DEIMOS, the staff of the Keck Observatory,
and the DEEP2 team for encouragement and assistance with observations
and data reduction. We acknowledge discussions with A. Benson,
B. Abraham, G. Bertin, R. Carlberg, K. Glazebrook, X. Hernandez,
M. Malkan, A. Renzini, S.White.  Based on data obtained with the HST
operated by AURA for NASA and with the W.M. Keck Observatory on Mauna
Kea, Hawaii, operated by Caltech, UC, and NASA and made possible by
the W.M.  Keck Foundation.  We thank the referee for constructive
criticism.

\begin{inlinefigure}
\begin{center}
\resizebox{\textwidth}{!}{\includegraphics{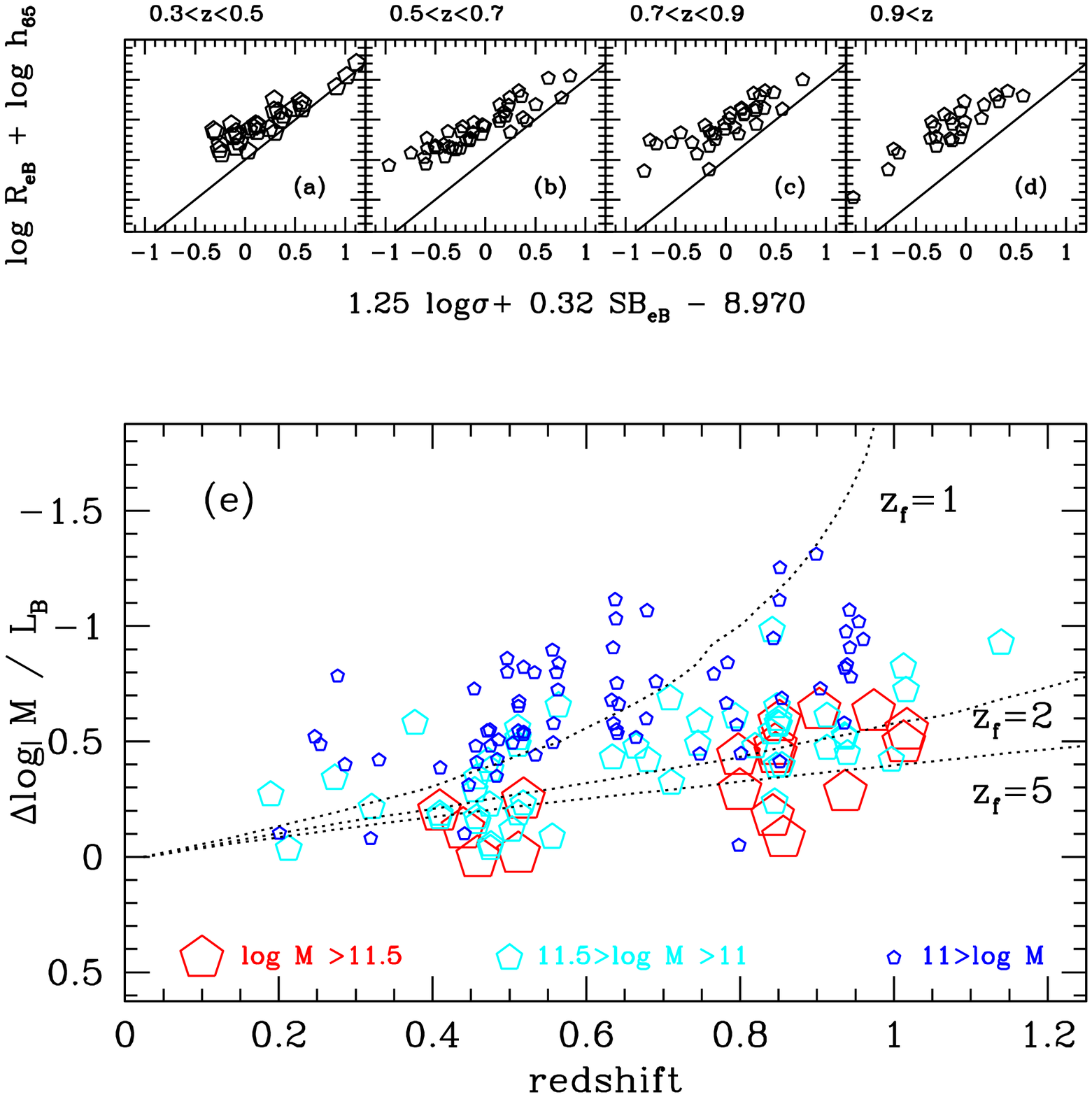}}
\end{center}
\figcaption{Top panels: edge-on view of the FP in redshift bins. The
solid line represents the FP for the Coma cluster.  Bottom panel:
offset from the Coma FP, coded by dynamical mass demonstrating the
stronger evolution for smaller masses. Typical errors are $<0.1$ dex
on the vertical axis. Dotted lines represent passive evolutionary
trends for single burst models at various formation epochs ($z_{\rm
f}$).
\label{fig:FPBp4}}
\end{inlinefigure}

\begin{inlinefigure}
\begin{center}
\resizebox{\textwidth}{!}{\includegraphics{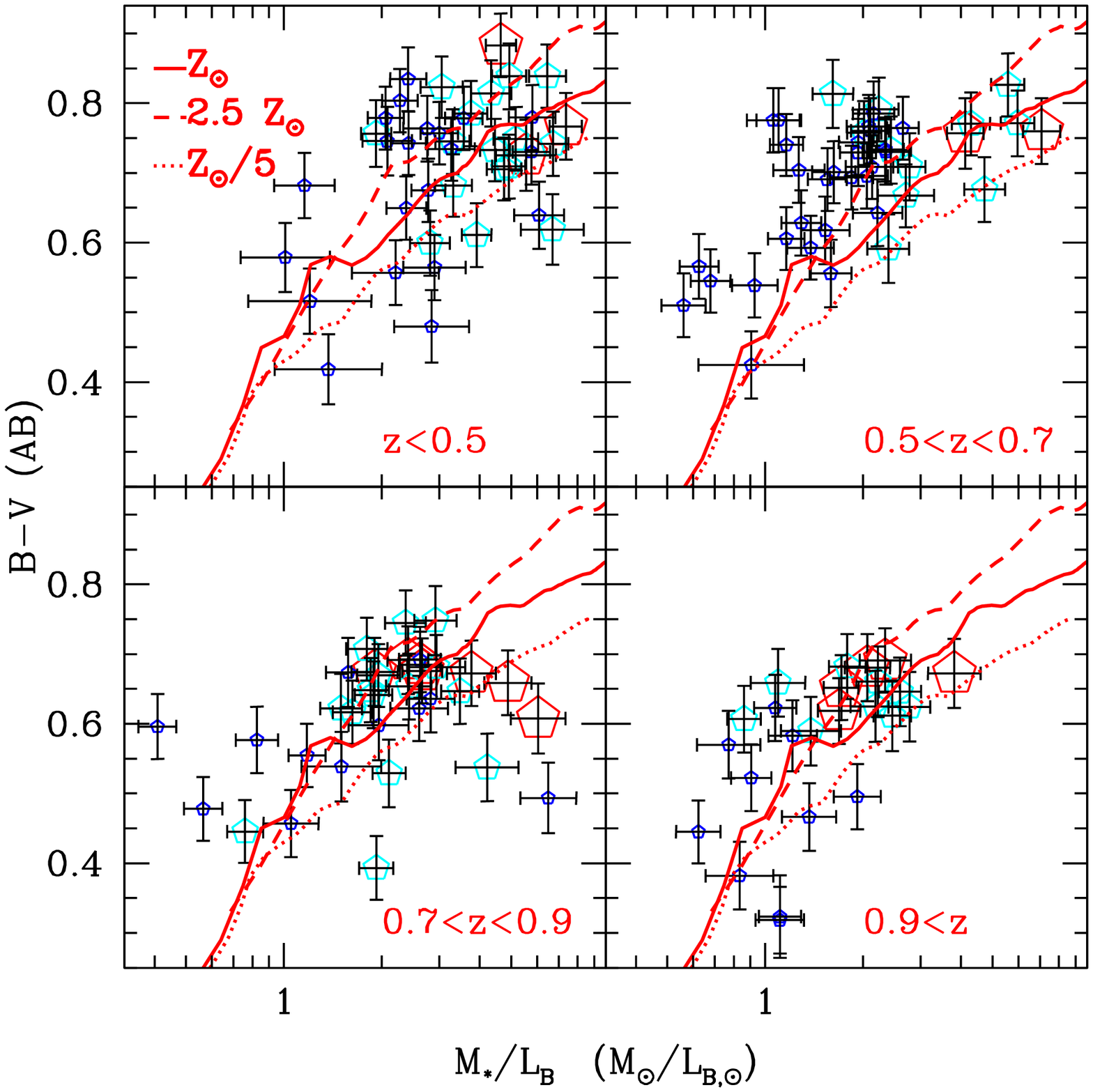}}
\end{center}
\figcaption{Rest frame (B-V) color as a function of the stellar mass
to light ratio, obtained assuming a local value of 7.3
$M_{\odot}/L_{B,\odot}$. The red curves are the tracks followed by
single burst stellar populations with Salpeter IMF (ages 0.5 - 12
Gyrs) and various metallicities. The size of symbols represents galaxy
mass as in Figure~1.
\label{fig:DFPBV}}
\end{inlinefigure}

\begin{inlinefigure}
\begin{center}
\resizebox{\textwidth}{!}{\includegraphics{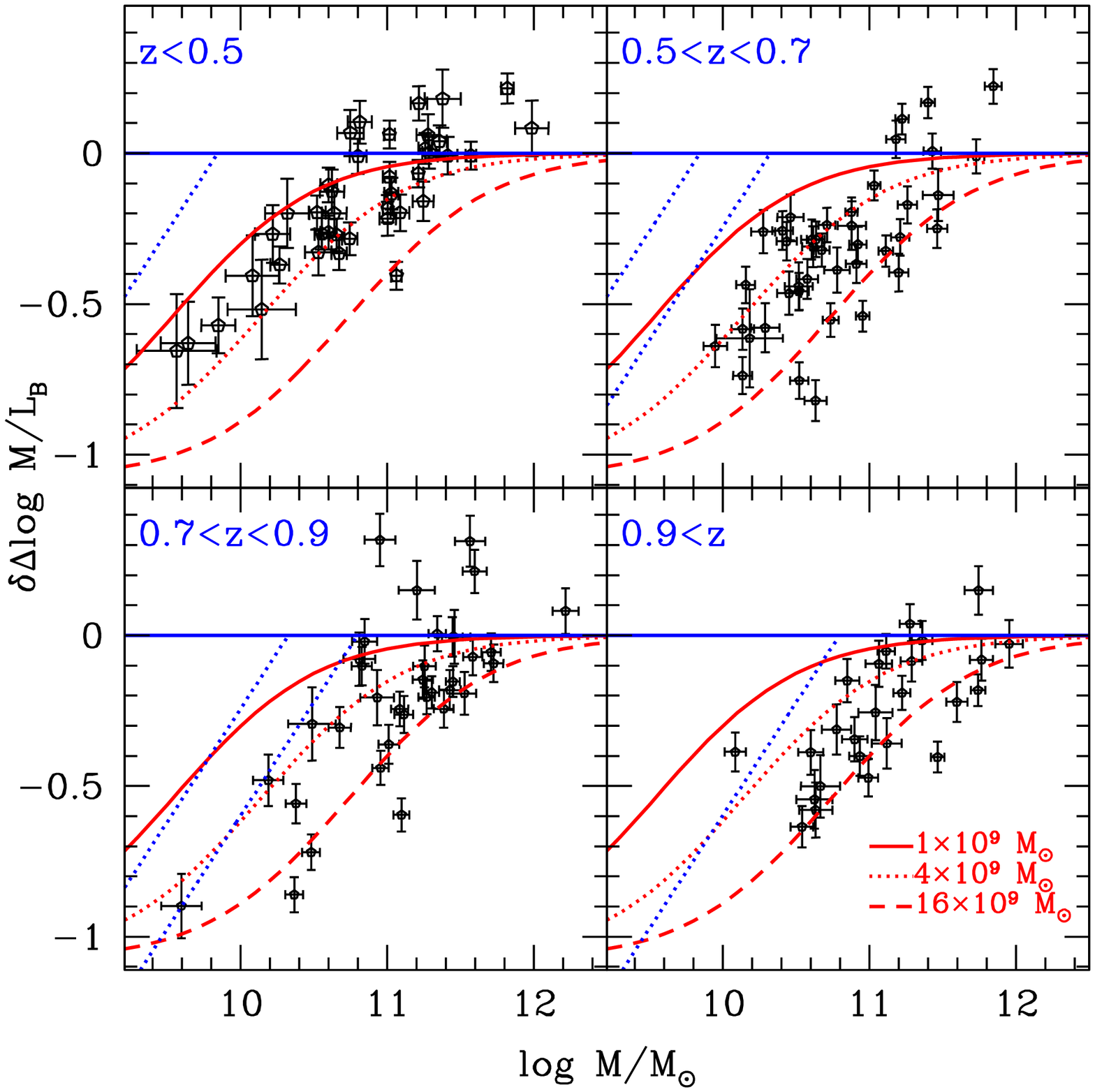}}
\end{center}
\figcaption{$M/L_{\rm B}$ offset with respect to the
redshift-dependent behavior of old massive cluster spheroidals,
plotted against dynamical mass $M$. The blue horizontal line
represents the locus seen for massive cluster galaxies.  Red curves
illustrate the effects of secondary activity (1 Gyr ago) applied to a
10 Gyr old population of mass $M$. The blue (dotted) diagonal lines
demonstrate the effect of the magnitude limit for each redshift bin.
\label{fig:DelDel4}}
\end{inlinefigure}

\begin{inlinefigure}
\begin{center}
\resizebox{\textwidth}{!}{\includegraphics{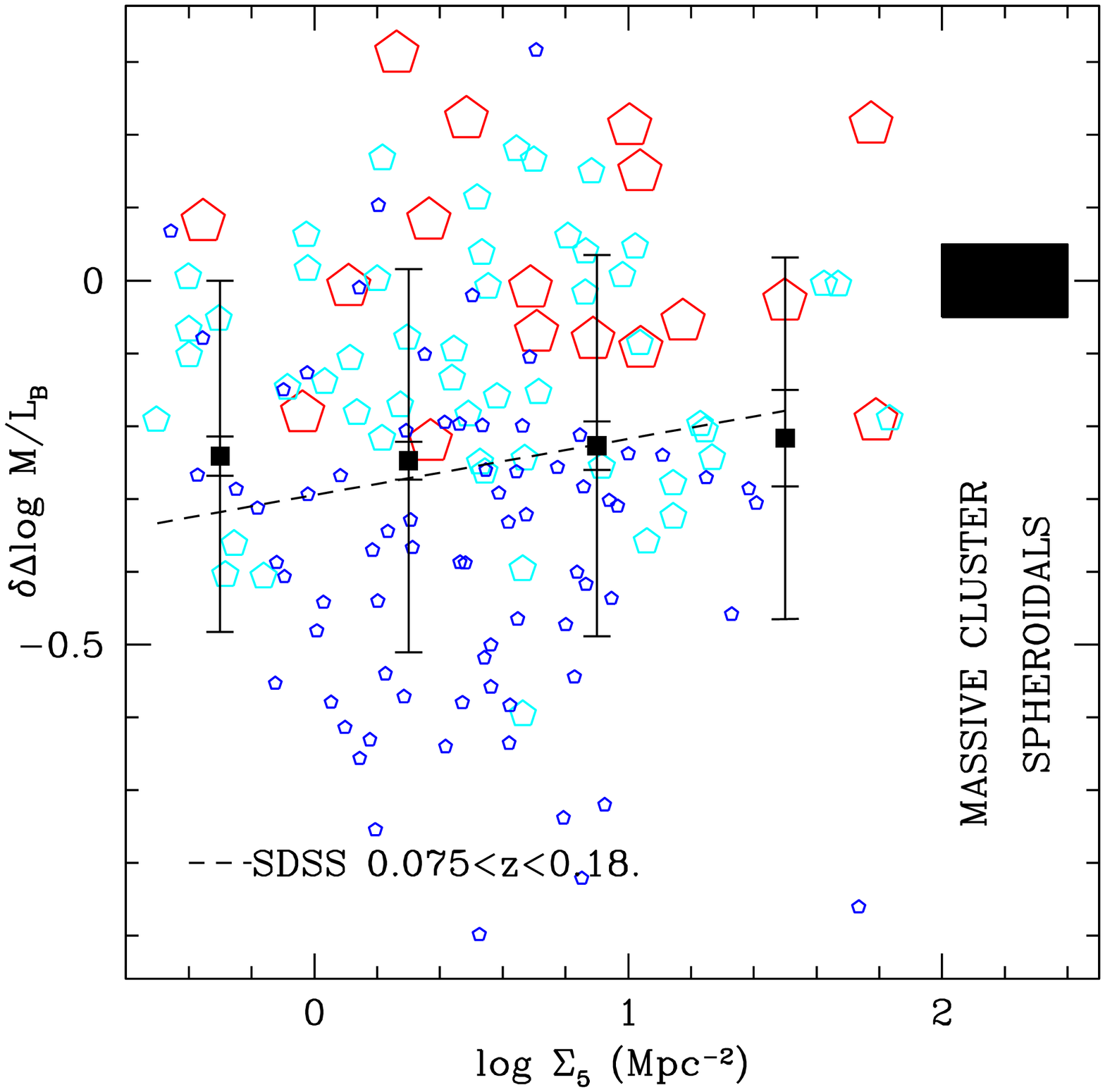}}
\end{center}
\figcaption{Offset of field spheroidals from the cluster trend
($\delta \Delta M/L_{\rm B})$ as a function of local projected density
$\Sigma_5$, coded by mass as in Figure~1 (The error on the local
density is $\sim 0.3$ dex). Solid squares represent the averages in
each density bin; large error bars represent the scatter, small error
bars the uncertainty in the mean. The black rectangle represents the
value observed for massive cluster spheroidals, the diagonal dashed
line the local trend from the SDSS Survey (Bernardi et al.\ 2003). 
\label{fig:DDS}}
\end{inlinefigure}

\end{document}